\documentclass[pra, showpacs, twocolumn, floatfix]{revtex4}
\usepackage{graphicx}
\usepackage{amsmath, amsfonts, amssymb, bm}
\usepackage[]{psfrag}
\psfragscanoff 

\newcommand{\vect}[1]{\boldsymbol{#1}}
\usepackage{braket}
\usepackage{natbib}
\usepackage{units}
\begin{document}
\title{Flexible generation of correlated photon pairs in different frequency ranges}
\author{Fernando \surname{Oster}}

\author{Christoph H. \surname{Keitel}}

\author{Mihai \surname{Macovei}}
\email{mihai.macovei@mpi-hd.mpg.de}

\affiliation{Max-Planck-Institut f\"{u}r Kernphysik, Saupfercheckweg 1, D-69117 Heidelberg, Germany}
\date{\today}
\begin{abstract}
The feasibility to generate correlated photon pairs at variable frequencies is investigated. 
For this purpose, we consider the interaction of an off-resonant laser field with a two-level system possessing 
broken inversion symmetry. We show that the system generates non-classical photon pairs exhibiting strong 
intensity-intensity correlations. The intensity of the applied laser tunes the degree of correlation while 
the detuning controls the frequency of one of the photons which can be in the THz-domain. Furthermore, we 
observe the violation of a Cauchy-Schwarz inequality characterizing these photons. 
\end{abstract}
\pacs{42.50.Dv, 42.50.Fx, 78.67.De, 78.67.Hc}
\maketitle
\section{Introduction}
The nature of light has always intrigued mankind and its study has nowadays culminated in the field of quantum 
optics investigating matter-field interaction \cite{scully,walls}. With the first measurement of an intensity-intensity correlation function by Hanbury Brown and Twiss \cite{hanbury} and the theoretical basis 
for the characterization of light by Glauber \cite{glauber,glauber2}, scientists have efficient tools in their 
hands to probe light fields for quantum signatures \cite{loudon}. In the last few decades, the interest in 
non-classical light has grown significantly with the advent of quantum computation and information science 
\cite{chuang}. Entangled photon pairs turn out to be indispensable for many quantum protocols \cite{horodecki} 
and quantum algorithms \cite{cleve}. Currently, there are a series of experimental techniques available to 
produce entangled photons such as parametric down conversion \cite{klyshko,pan,gagik}, four-wave mixing 
\cite{du,macovei,kiffner}, electromagnetically induced transparency \cite{balic,pasp} or cavity QED \cite{beige,zhu}. Furthermore, an atomic memory for correlated photon states has been realized experimentally, playing an essential role for quantum communication over long distances \cite{kuzmich,edamatsu,vdwal}. Recently, a heralded entanglement source of great practical importance has been demonstrated \cite{wagenknecht,barz}. In addition, theoretical considerations have predicted the generation of a correlated photon pair in the x-ray regime from strongly driven atomic ensembles \cite{jin}. 
Very recently, a communication network for quantum information processing has been proposed \cite{kimble}, which 
consists of numerous different nodes and channels. Since such different nodes may have different characteristic 
frequencies, there is great interest in investigating non-classical pairs of photons of different frequencies 
\cite{guo}. As an important milestone in this direction, entangled photons of different but close frequencies limited to the microwave or optical ranges have been generated and detected experimentally \cite{eichler,opt}. 

\begin{figure}[b]
\centering
\includegraphics[width=8cm]{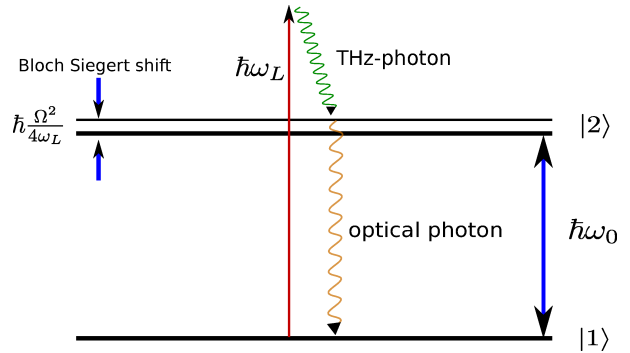}
\caption{(color online) The emission of the non-classical photon pair. The non-resonant laser excites the 
two-level system with broken inversion symmetry and induces the emission of a THz-photon and the subsequent 
spontaneously emitted optical photon.}
\label{fig:summary}
\end{figure}

Based on this background, we investigate here a two-level system with broken inversion symmetry which is driven by an off-resonant laser field. By means of adjusting the laser frequency $\omega_{L}$, one can spontaneously generate a photon at an approximate frequency $\omega_{L} -\omega_{0}$ and a subsequent photon with transition frequency $\omega_{0}$. With the parameters of e.g. gamma globulin macromolecules, those frequencies can be in the THz- and optical regimes, respectively, see Fig.~\ref{fig:summary}. We find that this photon-pair of different frequencies is both of non-classical character and entangled because it violates a Cauchy-Schwarz inequality. The advantage of our scheme lies in the fact that the frequency of the longer wavelength photon can be manipulated by an appropriate selected detuning. This is quite useful in driving a quantum network composed of different nodes of various frequencies including quantum wells or dots of THz transition frequencies. Furthermore, the high flexibility distinguishes our model from a cascade three-level system or other down conversion processes.

\section{The Model}
In particular, we consider a two-level system (see Fig.~\ref{fig:summary}) with the transition frequency 
$\omega_0$ described by the orthonormal ground state $\ket{1}$ and excited state $\ket{2}$ with broken 
inversion symmetry, meaning that the diagonal parts of the dipole operator satisfy the following condition: 
$|\vect{\wp}_{11}|\neq |\vect{\wp}_{22}|$, where we define $\vect{\wp}_{ij}=e\braket{i|\vect{r}|j}$ for $\{i,j\} \in \{1,2\}$. The system is driven by a classical off-resonant laser field given by a linearly polarized monochromatic plane-wave field in the dipole approximation ${\vect{E}=\vect{E}_0 \cos (\omega_L t)}$ with laser frequency $\omega_L$ and amplitude $\vect{E}_{0}$. The sample is surrounded by a quantized environment that accounts for the processes of spontaneous emission \cite{rew_mart}. 

The Hamiltonian $H$ describing the system takes into account the energy of the environment $H_E$ and of the two-level system $H_T$, the interaction-energy between the laser and the two-level system $H_{I1}$ and the interaction-energy between the environment and the two-level system $H_{I2}$: $H=H_E+H_T+H_{I1}+H_{I2}$, or
\begin{eqnarray}
H=\sum_{\vect{k}}\hbar \omega_{\vect{k}} a_{\vect{k}}^\dagger a_{\vect{k}} + \hbar \omega_{0}S_{z} 
+\hbar \Omega(S^{+} + S^{-})\cos(\omega_{L}t) \nonumber \\
+ \hbar G S_{z}\cos(\omega_{L}t ) + i \sum_{\vect{k}} (\vect{g}_{\vect{k}} \cdot \vect{d})
(a_{\vect{k}}^\dagger - a_{\vect{k}})(S^{+} + S^{-}), \label{eqn:initialhamiltonian}
\end{eqnarray}
where we define the Rabi frequency $\Omega=\vect{\wp}_{12} \cdot \vect{E}_0/\hbar$, and 
$G=(\vect{\wp}_{11}-\vect{\wp}_{22}) \cdot \vect{E}_0/\hbar$ leads to broken inversion 
symmetry \cite{bis}. Here, we have introduced the usual atomic 
operators $S^{+}=\ket{2}\bra{1}$, $S^{-}=\ket{1}\bra{2}$ and $S_{z}=(\ket{2}\bra{2}-\ket{1}\bra{1})/2$. $a_{\vect{k}}^\dagger$ and $a_{\vect{k}}$ denote the creation and annihilation electromagnetic field 
operators of the $\vect{k}$-th mode of frequency $\omega_{\vect{k}}$. The coupling constant $\vect{g}_{\vect{k}}$ is defined as 
$\vect{g}_{\vect{k}}=\sqrt{2\pi\hbar\omega_{\vect{k}}/V}\hat{\epsilon}_{\lambda}$, where 
$\hat{\epsilon}_{\lambda}$ is the photon polarization vector, $\lambda \in \{1,2\}$, and $V$ is the electromagnetic field quantization volume.
The electromagnetic atom-field interaction is given in the usual dipole-approximation. We stress the fact 
that we do not work in the rotating wave approximation, but rather choose a perturbative approach to 
account for non-linear effects. For this purpose, we first perform a unitary transformation on $H$ with 
$H_0=\sum_{\vect{k}}\hbar \omega_{L}a_{\vect{k}}^\dagger a_{\vect{k}} + \hbar \omega_{L}S_{z}$,
\begin{eqnarray}
\label{eqn:rotframe}
	\tilde{H}=e^{\frac{i}{\hbar} H_0 t}(H-H_0) e^{-\frac{i}{\hbar} H_0 t},
\end{eqnarray}
which may be separated ($\tilde{H}=\tilde{H}'+\tilde{H}''$) into a time-independent part
\begin{eqnarray}
\label{eqn:rwa}
&{}&\tilde{H}'=\sum_{\vect{k}} \hbar(\omega_{\vect{k}}-\omega_L) a_{\vect{k}}^\dagger a_{\vect{k}} 
+ \hbar(\omega_{0}-\omega_{L}) S_{z} \nonumber \\
&+& \frac{\hbar \Omega }{2}(S^{+} + S^{-} ) + 
i\sum_{\vect{k}}(\vect{g}_{\vect{k}} \cdot \vect{d})(a_{\vect{k}}^\dagger S^{-} - a_{\vect{k}} S^{+}),
\end{eqnarray}
and a time-dependent part containing fast oscillating terms
\begin{eqnarray}
\label{eqn:nrwa}
\tilde{H}''&=& \frac{\hbar G}{2}S_z( e^{i \omega_L t} + e^{-i \omega_L t}) \nonumber \\
&+& \frac{\hbar \Omega }{2}(S^+ e^{2i \omega_L t}+S^- e^{-2i \omega_L t})  \nonumber \\
&+& i \sum_{\vect{k}} (\vect{g}_{\vect{k}} \cdot \vect{d}) (a_{\vect{k}}^{\dagger}S^{+}e^{2i\omega_L t}
-a_{\vect{k}} S^{-}e^{-2i \omega_L t}).
\end{eqnarray}
The time-dependent part can be regarded as a perturbation to the time-independent part and we can thus 
apply the second-order perturbation theory \cite{ficek,james}, since $G < \omega_L$, $\Omega < \omega_L$ 
and $(\vect{g}_{\vect{k}} \cdot \vect{d}) < \omega_L$:
\begin{eqnarray}
\label{eqn:perturbationtheory}
H_{pert} =-\frac{i}{\hbar}\tilde{H}''\int dt \tilde{H}''.
\end{eqnarray}
Our final Hamiltonian $H_{f}=\tilde{H}'+H_{pert}$ acquires the shape
\begin{eqnarray}
\label{eqn:finalhamiltonian}
H_{f}&=&\sum_{\vect{k}} \hbar(\omega_{\vect{k}}-\omega_L) a_{\vect{k}}^\dagger a_{\vect{k}} 
+ \hbar(\omega_{0} - \omega_{L} + \frac{\Omega^2 }{4\omega_L})S_{z} \nonumber \\
&+&\frac{\hbar \Omega }{2}(S^{+} + S^{-}) + i\sum_{\vect{k}}(\vect{g}_{\vect{k}} \cdot \vect{d})
(a_{\vect{k}}^\dagger S^{-} - a_{\vect{k}} S^{+}) \nonumber \\
&+& \frac{3G}{8i\omega_L}\sum_{\vect{k}}(\vect{g}_{\vect{k}} \cdot \vect{d})
(a_{\vect{k}}^\dagger S^{+} e^{i\omega_{L} t} - a_{\vect{k}} S^{-}e^{-i\omega_L t}) \nonumber \\
&+&\frac{\Omega}{2i\omega_L} \sum_{\vect{k}} 
(\vect{g}_{\vect{k}} \cdot \vect{d})(a_{\vect{k}}-a_{\vect{k}}^\dagger)S_{z},
\end{eqnarray}
where we keep the slowlyest oscillating time-dependent terms only. We notice that the time-dependent terms 
are proportional to $G$ and are thus important for the description of a system with broken inversion symmetry. 
The ratios $G/\omega_{L}$ and $\Omega/\omega_{L}$ are small for optical frequencies $\omega_{L}$ such that higher orders are negligible in the Hamiltonian. Our perturbative approach also reveals an effect of strong driving fields - the Bloch-Siegert shift  $\hbar\Omega^2/(4\omega_L)$ \cite{blochsiegert} of the upper state of the two-level system, see Fig.~\ref{fig:summary}. Finally, the two-level approximation applies because 
$\Omega/\omega_{L} \ll 1$ and $|\omega_{0}-\omega_{L}|/\omega_{L} \ll 1$.

In what follows, we shall derive the master equation employing the Hamiltonian in 
Eq.~(\ref{eqn:finalhamiltonian}) and the Heisenberg picture. We assume that the matter-field 
interaction is weak in the sense that an emitted photon does not react back on the atom and use the well-known Born-Markov approximation. Thus, the time-evolution of an arbitrary atomic operator $Q(t)$ is governed by the Heisenberg equation: $ \frac{d}{dt} \braket{Q(t)}=\frac{i}{\hbar} \braket{[H_{f},Q]}.$
By inserting the final Hamiltonian, we obtain
\begin{eqnarray}
\label{eqn:masterequation}
&{}&\frac{d}{dt}\braket{Q(t)}=\frac{i}{\hbar} \braket{[\tilde{H_0},Q]} \nonumber \\
&-&\sum_{\vect{k}} \frac{(\vect{g}_{\vect{k}} \cdot \vect{d})}{\hbar} \{ \braket{a_{\vect{k}}^\dagger[S^-,Q]}+\braket{[Q,S^+]a_{\vect{k}}} \nonumber \\
&-& \frac{3G}{8\omega_L}(\braket{a_{\vect{k}}^\dagger[S^+,Q]}e^{i\omega_L t} 
+ \braket{[Q,S^-]a_{\vect{k}}}e^{-i\omega_L t}) \nonumber \\
&+& \frac{\Omega}{2 \omega_L}(\braket{a_{\vect{k}}^\dagger[S_z,Q]}+\braket{[Q,S_z]a_{\vect{k}}})\},
\end{eqnarray}
where $\tilde{H_0}=\hbar(\omega_0-\omega_L+\Omega^2 /(4\omega_L)) S_z+\hbar \Omega(S^+ + S^- )/2.$
To further simplify the analytical formalism, we have to express the annihilation and creation operators 
as a function of atomic operators in the Born-Markov approximation. First, we insert 
$a_{\vect{k}}^\dagger(t)$ in the Heisenberg equation and obtain the general solution for the linear inhomogeneous differential equation of first order. Then we consider the leading order in the coupling 
and neglect the Lamb-shift, so that the creation operator acquires the shape
\begin{eqnarray}
\label{eqn:creationoperator}
&{}&a_{\vect{k}}^\dagger(t)=a_{\vect{k}}^\dagger(0) e^{i\Delta_{\vect{k}}t} 
+ \frac{\pi\Omega}{2\hbar \omega_L} (\vect{g}_{\vect{k}} \cdot \vect{d}) S_z(t)\delta(\Delta_{\vect{k}}) \nonumber \\
&-&\frac{3\pi G}{8\hbar \omega_L} (\vect{g}_{\vect{k}} \cdot \vect{d})S^-(t)
\delta(\Delta_{\vect{k}} +\omega_0+\frac{\Omega^2}{4\omega_L}) e^{-i\omega_L t} \nonumber \\
&+& \pi \frac{(\vect{g}_{\vect{k}} \cdot \vect{d})}{\hbar} S^+(t) 
\delta(\omega_{\vect{k}}-\omega_0-\frac{\Omega^2}{4 \omega_L}),
\end{eqnarray}
where $\Delta_{\vect{k}}=\omega_{\vect{k}}-\omega_L$. 
We notice that for the annihilation operator $a_{\vect{k}}$, we only have to take the H.c. of the above formula. If we further define the different decay rates of the system
\begin{subequations}\label{eqn:decayrates}
\begin{align}
\gamma_R&=\pi \sum_{\vect{k}} \frac{(\vect{g}_{\vect{k}} \cdot \vect{d})^2}{\hbar^2} \delta(\omega_{\vect{k}}-\omega_0-\frac{\Omega^2}{4 \omega_L}),\\
\gamma_L&=\pi \sum_{\vect{k}} \frac{(\vect{g}_{\vect{k}} \cdot \vect{d})^2}{\hbar^2} \delta(\omega_{\vect{k}}-\omega_L),\\
\gamma_T&	=\pi \sum_{\vect{k}}  \frac{(\vect{g}_{\vect{k}} \cdot \vect{d})^2}{\hbar^2} \delta(\omega_{\vect{k}}-\omega_L+\omega_0+\frac{\Omega^2}{4 \omega_L}),
\end{align}
\end{subequations}
and insert Eq.~(\ref{eqn:creationoperator}) in Eq.~(\ref{eqn:masterequation}), we may write down our final master equation
\begin{eqnarray}
\label{eqn:equationofmotion}
&{}&\frac{d}{dt} \braket{Q(t)}-\frac{i}{\hbar} \braket{[\tilde H_0,Q]}\nonumber \\
&=& -\gamma_R(\braket{S^+[S^-,Q]}+\braket{[Q,S^+]S^-})\nonumber \\
&-&\frac{\Omega}{2 \omega_L} \gamma_L(\braket{S_z[S^-,Q]}+\braket{[Q,S^+]S_z})\nonumber \\
&-& (\frac{3G}{8\omega_L})^2 \gamma_T(\braket{S^-[S^+,Q]}+\braket{[Q,S^-]S^+})\nonumber \\
&-& \frac{\Omega}{2 \omega_L} \gamma_R(\braket{S^+[S_z,Q]} +\braket{[Q,S_z]S^-} )\nonumber \\
&-& (\frac{\Omega}{2 \omega_L})^2 \gamma_L(\braket{S_z[S_z,Q]}+\braket{[Q,S_z]S_z}),
\end{eqnarray}
which may be interpreted as follows: the first term accounts for the spontaneous emission at resonance 
$\omega_{0} + \Omega^{2}/(4 \omega_{L})$, taking into account the Bloch-Siegert shift. The second term 
describes the spontaneous emission at the laser frequency $\omega_{L}$ preceded by an excitation. 
The third term corresponds to the emission at frequency $\omega_{L} - \omega_{0} - \Omega^{2}/(4 \omega_{L})$ preceded by an excitation of the two-level system. With the used parameters, later on, it has THz-frequency 
while the main resonance is optical. The fourth term accounts for a spontaneous emission at resonance preceded by an excitation (off-resonant as always). The last term contributes to the dephasing of the system. We are interested in correlations between the 
processes of the first and third summands that are illustrated in Fig.~\ref{fig:summary}. For this purpose, 
we need to define these correlations and their time-dependent behaviors. 

In order to probe the quantum nature of our generated photons, we calculate its intensity-intensity 
correlation function $g_{ij}^{(2)}$ defined as \cite{scully,walls}
\begin{gather}
g_{ij}^{(2)}(\tau)=\frac{\langle E_i^{(-)}(t)E_j^{(-)}(t+\tau)E_j^{(+)}(t+\tau)E_i^{(+)}(t) \rangle}{\langle E_i^{(-)}(t)E_i^{(+)}(t) \rangle \langle E_j^{(-)}(t)E_j^{(+)}(t) \rangle}.
\end{gather}
We know from the definition of the quantized electric field \cite{scully} that $\vect{E}^{(-)} \propto a_{\vect{k}}^{\dagger}$ and that $\vect{E}^{(+)} \propto a_{\vect{k}}$. In our case, we also know from Eq.~(\ref{eqn:creationoperator}) that for THz-emission $a_{\vect{k}}^{\dagger} \propto S^-$ and that for 
optical emission ${a_{\vect{k}}^{\dagger} \propto S^+}$. Therefore, the probability for detecting 
an optical photon after a THz-photon as a function of atomic operators is given by
\begin{eqnarray}
g_{12}^{(2)}(0)=\frac{\langle S^-(t)S^+(t)S^-(t)S^+(t)\rangle}{\langle S^-(t)S^+(t)\rangle 
\langle S^+(t)S^-(t)\rangle}
\end{eqnarray}
and the probability for detecting an optical photon followed by a THz-photon reads
\begin{eqnarray}
g_{21}^{(2)}(0)=\frac{\langle S^+(t)S^-(t)S^+(t)S^-(t)\rangle}{\langle S^+(t)S^-(t)\rangle 
\langle S^-(t)S^+(t)\rangle}.
\end{eqnarray}
\begin{figure}[t]
\centering
\includegraphics[width=8cm]{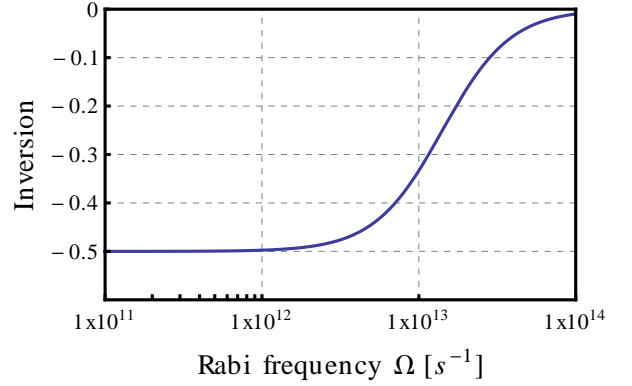}
\caption{The steady-state inversion operator as a function of the Rabi frequency 
with transition frequency $\omega_0 = \unit[5.0 \times 10^{15}]{s^{-1}}$, 
laser frequency $\omega_L=\omega_0+\unit[10^{13}]{s^{-1}}$, detuning 
$\Delta=\unit[10^{13}]{s^{-1}}$ and decay rate $\gamma_0=\unit[3 \times 10^{6}]{s^{-1}}$ 
with respect to $\omega_0$.}
\label{fig:population}
\end{figure}

As a concrete system, we consider gamma globulin macromolecules \cite{kovarskii} with the following parameters  ${|\omega_2-\omega_1| \cong \unit[4.8 \times 10^{15}]{s^{-1}}}$, $|\vect{\wp}_{21}|\cong \unit[1]{D}$ and ${|\vect{\wp}_{22}-\vect{\wp}_{11}|\cong \unit[100]{D}}$. We notice that the transition frequency is optical and 
we do observe the necessary broken inversion symmetry. We choose the laser detuning such that the long wavelength photon is in the THz domain. Alternative systems are quantum dots, which are 0-dimensional quantum systems having an electron confined in all three space dimensions \cite{kittel}. Gallium nitride devices for example show broken inversion symmetry and have typical values of  ${|\vect{\wp}_{22}-\vect{\wp}_{11}|\cong \unit[10]{D}}$, $|\vect{\wp}_{12}|\cong \unit[10]{D}$ and ${\omega_0=\unit[4.92 \times 10^{15}]{s^{-1}}}$ \cite{kibis,williams,bretagnon,rinke}.

\section{Results}
\begin{figure}[t]
\centering
\begin{minipage}{0.5\textwidth}
\includegraphics[width=6.2cm]{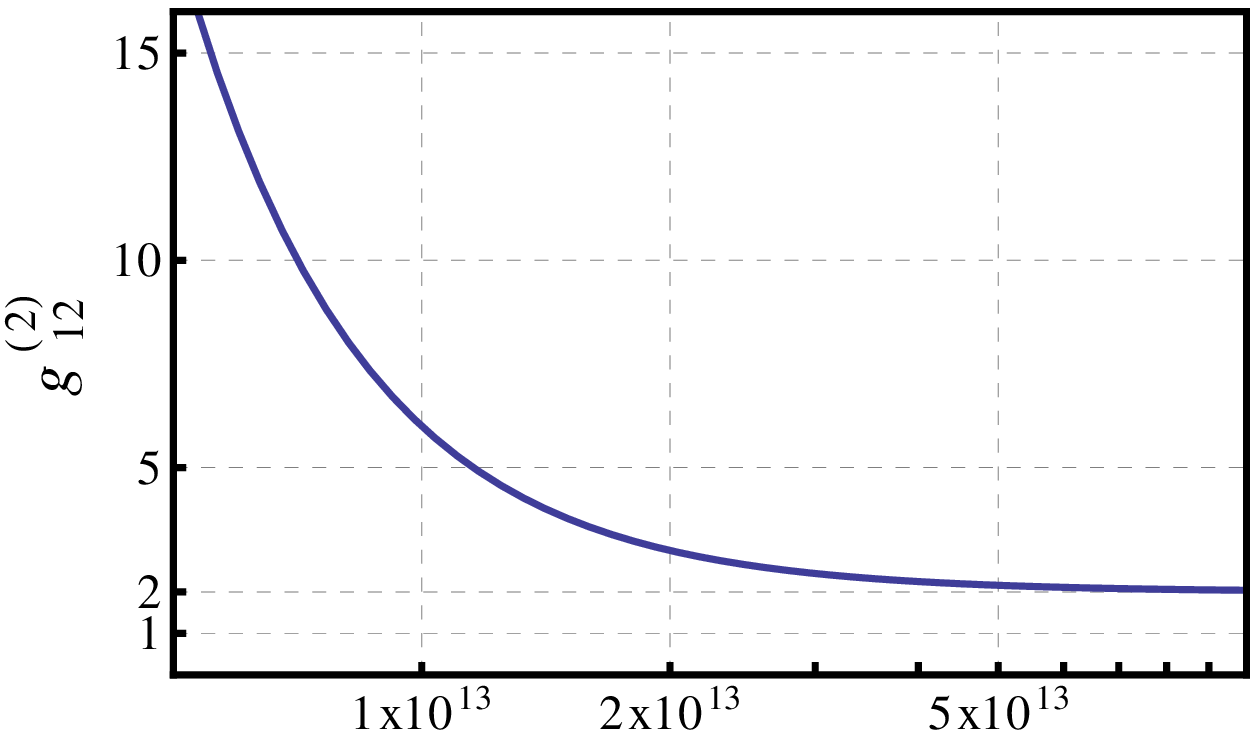}
\end{minipage}
\\\vskip-0.3cm
\begin{minipage}{0.5\textwidth}
\includegraphics[width=6.6cm]{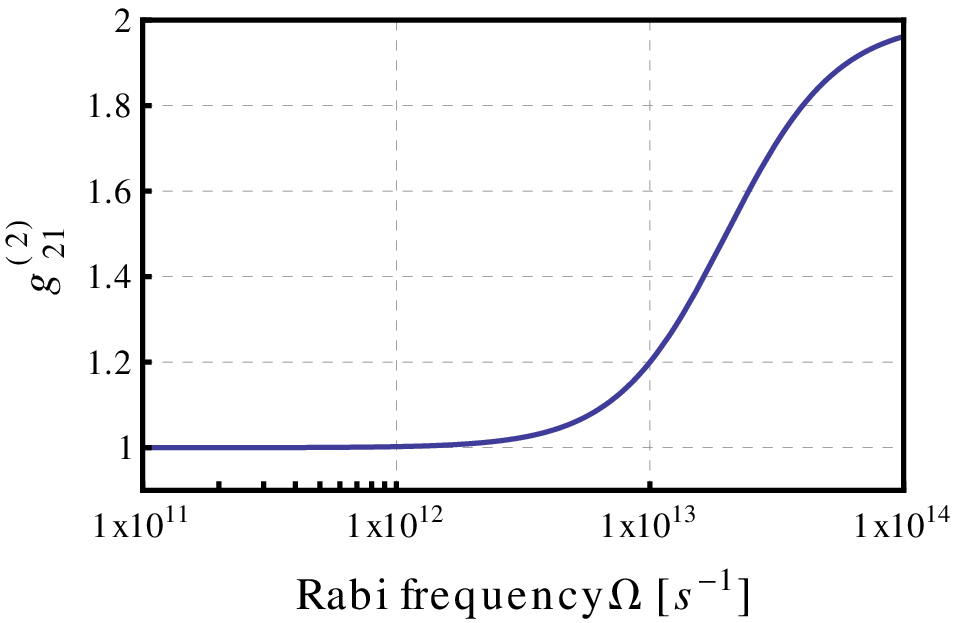}
\end{minipage}
\begin{picture}(0,0)
 \put(-40,215){(a)}
\put(-40,108){(b)}
\end{picture}

\caption{The steady-state second-order intensity-intensity correlation function describing the 
probability of (a) THz-emission followed by an optical emission and (b) optical emission followed by 
THz-emission as a function of the Rabi frequency $\Omega$. Otherwise, we use the same parameters as in Fig.~\ref{fig:population}.}
\label{fig:correl}
\end{figure}

At first, we display the population inversion $\braket{S_z(t)}$ as a function of the Rabi frequency $\Omega$ in 
Fig.~\ref{fig:population}. We observe that for low Rabi frequencies $\Omega$, the population remains in the 
ground state. At a frequency of about $\unit[10^{12}]{s^{-1}}$, we notice an increase of the population and 
at $\unit[10^{13}]{s^{-1}}$, we see that there is a non-vanishing probability to find the system in the excited 
state. Now, we turn to the plot in Fig.~\ref{fig:correl}(a) of the second-order correlation function $g_{12}^{(2)}(0)$ as a function of the Rabi frequency $\Omega$ describing the probability of the emission of a THz-photon and the subsequent emission of an optical photon. We observe a strong correlation which decreases with rising Rabi frequency. To induce the emission of a THz-photon, the system has to be excited from the ground state to the upper state, where it may spontaneously emit an optical photon. Thus, at low Rabi frequencies, the 
emission of an optical photon is almost always preceded by the emission of a THz-photon. This explains the 
high degree of correlation of the photon pair. As $\Omega$ increases, there is a non-vanishing probability to 
find the system in the excited state and an optical emission that is not preceded by a THz-photon is possible. 
This means that the correlation decreases. Finally, we discuss the intensity-intensity correlation function 
$g_{21}^{(2)}(0)$ in Fig.~\ref{fig:correl}(b) describing the probability of detecting a THz-photon right after 
an optical photon. It turns out that this probability is very low as expected. It slowly rises with increasing 
Rabi frequency $\Omega$. 

In this context, we also investigate the violation of the Cauchy-Schwarz inequality
\begin{eqnarray}\label{eqn:cauchyschwarz}
g_{11}^{(2)}(0)g_{22}^{(2)}(0) \geq [g_{12}^{(2)}(0)]^2.
\end{eqnarray}
The correlations $g_{11}^{(2)}(0)$ and $g_{22}^{(2)}(0)$ vanish trivially and in Fig.~\ref{fig:correl}(a,b), 
we notice nonvanishing cross-correlations violating Eq.~(\ref{eqn:cauchyschwarz}). Thus, we 
are dealing with a non-classical pair of correlated and entangled photons.

\section{Summary}
In summary, we have investigated the interaction of a two-level system with broken inversion symmetry and an 
off-resonant laser field. Using the parameters of e.g. gamma globulin macromolecules or certain quantum dots, we have found the possibility to generate a long wavelength photon in the THz-regime followed by a photon in 
the optical frequency range. Furthermore, we have observed a high degree of correlation between these photons and even a violation of a Cauchy-Schwarz inequality. This proves the non-classical character and entanglement 
of the photon pair. In the emerging field of quantum information science, non-classical correlated or even entangled photon pairs of different frequencies are of great interest, finding applications in the realization of a quantum network.


\end{document}